\newcommand{\ebr}{E$_{{\mbox{br}}}$}
\newcommand{\spe}{S$_{{\mbox{e}}}$}
\newcommand{\spl}{S$_{{\mbox{l}}}$}
\documentclass{aa}
\usepackage{graphicx}
\usepackage{txfonts}
\begin{document}

\title{The ESO Nearby Abell Cluster Survey
       \thanks{Based on observations collected at the European Southern
               Observatory (La Silla, Chile)}
       \thanks{http://www.astrsp-mrs.fr/www/enacs.html}
      }

\subtitle{XIII. The orbits of the different types of galaxies in 
                rich clusters}

\author{A.~Biviano \inst{1} \and P.~Katgert \inst{2}}
\institute{INAF/Osservatorio Astronomico di Trieste, Italy  
\and Sterrewacht Leiden, The Netherlands} 

\offprints{A. Biviano, \email{biviano@ts.astro.it}}

\date{Received / Accepted}

\markboth{ENACS XIII}{Galaxy orbits in clusters}

\abstract{We study the orbits of the various types of galaxies
observed in the ESO Nearby Abell Cluster Survey. We combine the
observed kinematics and projected distributions of galaxies of various
types with an estimate of the mass density profile of the ensemble
cluster to derive velocity-anisotropy profiles. Galaxies within and
outside substructures are considered separately. Among the galaxies
outside substructures we distinguish four classes, on the basis of
their projected phase-space distributions. These classes are: the
brightest ellipticals (with $M_R \leq -22+5 \log h$), the other
ellipticals together with the S0's, the early-type spirals (Sa--Sb),
and the late-type spirals and irregulars (Sbc-Irr) together with the
emission-line galaxies (except those of early morphology). The mass
profile was determined from the distribution and kinematics of the
early-type (i.e. elliptical and S0) galaxies outside substructures;
the latter were assumed to be on isotropic orbits, which is supported
by the shape of their velocity distribution. The projected
distribution and kinematics of the galaxies of other types are used to
search for equilibrium solutions in the gravitational potential
derived from the early-type galaxies. We apply the method described by
Binney \& Mamon as implemented by Solanes \& Salvador-Sol\'e to
derive, to our knowledge for the first time, the velocity
anisotropy profiles of all galaxy classes individually (except, of
course, the early-type class). We check the validity of the solutions
for $\beta'(r) \equiv [{\rm{<} v_r^2 \rm{>}(r)}/{\rm{<} v_t^2
\rm{>}(r)}]^{1/2} $, where $\rm{<} v_r^2 \rm{>}(r)$ and 
$\rm{<} v_t^2 \rm{>}(r)$ are the mean squared components of the radial
and tangential velocity, respectively, by comparing the observed and
predicted velocity-dispersion profiles. For the brightest ellipticals
we are not able to construct equilibrium solutions. This is most
likely the result of the formation history and the special location of
these galaxies at the centres of their clusters. For both the early
and the late spirals, as well as for the galaxies in substructures,
the data allow equilibrium solutions. The data for the early spirals
are consistent with isotropic orbits ($\beta'(r) \equiv 1$), although
there is an apparent radial anisotropy at $\simeq 0.45 \,
r_{200}$. For the late spirals an equilibrium solution with isotropic
orbits is rejected by the data at the $> 99$\% confidence level. While
$\beta'(r) \approx 1$ within $0.7 \, r_{200}$, $\beta'$ increases
linearly with radius to a value $\simeq 1.8$ at $1.5 \, r_{200}$.
Taken at face value, the data for the galaxies in substructures
indicate that isotropic solutions are not acceptable, and tangential
orbits are indicated. Even though the details of the tangential
anisotropy remain to be determined, the general conclusion appears
robust. We briefly discuss the possible implications of these
velocity-anisotropy profiles for current ideas of the evolution and
transformation of galaxies in clusters.
 }

\maketitle

\section{Introduction}
\label{s-intro}

The orbital characteristics of the various types of galaxies in
present-day clusters can give unique information about the evolution
of the clusters themselves, and about the formation and evolution of
their member galaxies. This is because clusters are still accreting
galaxies from their surroundings, and the details of this accretion
process provide constraints for theories of cluster evolution. In
addition, the orbits of the various types of galaxies yield clues
about the history of their accretion onto the cluster, and about the
evolutionary relationships between them.

The idea of shells of collapsing material around clusters has been
around since the work of Gunn \& Gott (\cite{gun72}). This work stimulated
several investigations, and indirect evidence for the infall of
spirals into clusters has been accumulating over the years. Moss \&
Dickens (\cite{mos77}) were the first to observe a difference in the velocity
dispersions of cluster early-type and late-type galaxies, followed by
Sodr\'e et al. (\cite{sod89}), and Biviano et al. (\cite{biv92}). The different
projected phase-space distributions of early-type (red) and late-type
(blue) galaxies was clearly established by Colless
\& Dunn (\cite{col96}) and Biviano et al. (\cite{biv96}) in the Coma cluster,
while Carlberg et al. (\cite{car97a}) found it in the CNOC clusters at $z
\approx 0.3$. The effect was studied in detail for the clusters observed in
the ESO Nearby Abell Cluster Survey (ENACS, hereafter; de Theije \&
Katgert \cite{det99}, paper VI; Biviano et al. \cite{bivk02}, paper
XI), as well as in other clusters (see, e.g., Adami et
al. \cite{ada98a}). Galaxies with emission lines (ELG) provide a rather
extreme example of the effect. The ELG are less centrally concentrated
and have a higher dispersion of line-of-sight velocity than the
galaxies without emission lines. This was first shown by Mohr et
al. (\cite{moh96}) for the A576 cluster, and clearly demonstrated by Biviano
et al. (\cite{biv97}, paper III) for the ENACS clusters.

These results suggest mildly radial orbits of the late-type galaxies
with emission lines, probably in combination with first approach to
the central dense core. This interpretation would be consistent with
the presence of the line-emitting gas which is unlikely to `survive'
when the galaxy crosses the cluster core.  Indeed, Pryor \& Geller
(\cite{pry84}) tried to constrain the orbits of HI-deficient galaxies
by noting that cluster-core crossing is a necessary condition for gas
stripping, and Solanes et al. (\cite{sol01}) noted that the
velocity-dispersion profile of HI-deficient galaxies is quite steep,
suggestive of radial orbits. Support for the scenario of spiral infall
into clusters comes from the the analyses of the Tully-Fisher
distance-velocity diagram (Tully \& Shaya \cite{tul84}; Gavazzi et al.
\cite{gav91}). Indirect support comes from the numerical simulations 
that show that dark matter particles have a moderate radial velocity
anisotropy, which increases out to the virial radius (e.g. Tormen et
al. \cite{tor97}; Ghigna et al. \cite{ghi98}; Diaferio
\cite{dia99}). Radio or X-ray trails of cluster galaxies can also be
used to constrain their orbits (Merrifield \cite{mer98}).

In the absence of full dynamical modelling, the analysis of the galaxy
spatial distribution and kinematics can only suggest, but not really
constrain, the nature of cluster galaxy orbits. This is because the
projected spatial distribution, kinematics and mass model are coupled.
So far, only a few full dynamical analyses of the orbital distribution
of cluster galaxies exist. One reason for this is the relative paucity
of detailed data on the kinematics and distributions of cluster
galaxies, in particular if several galaxy classes are considered.
Another reason is that the orbital characteristics can only be
inferred from the observed kinematics and distributions if the mass
density profile of the cluster is known. The latter must be derived
either from the distribution of light (with assumptions about the
radial variation of the mass-to-light ratio), or from the projected
phase-space distribution of {\em that} subset of the galaxies for
which the properties of their full phase-space distribution can be
estimated independently.

A first dynamical analysis of the orbits of cluster galaxies was made
for the Coma cluster by Kent \& Gunn (\cite{ken82}). Using several analytical
mass models, these authors concluded that the galaxy orbits in the
Coma cluster cannot be primarily radial, so that even at large radii a
significant part of the kinetic energy of the galaxies must be in the
tangential direction. They noted that the range of the predicted
velocity dispersions of the galaxies of different morphological types
was only half that which is observed. Although a marginal result, this
could indicate different distribution functions for the galaxies of
different types, and not just different energy distributions. Merritt
(\cite{mer87}) used the same data to estimate the orbital anisotropy of the
galaxies in the Coma cluster, for various assumptions about the radial
dependence of the mass-to-light ratio.

More recent dynamical modelling of galaxy clusters has led to the
conclusion that the orbits of early-type galaxies are quasi-isotropic,
while those of late-type galaxies are moderately radial (e.g.
Natarajan \& Kneib \cite{nat96}; Carlberg et al. \cite{car97b};
Mahdavi et al. \cite{mah99}; Biviano \cite{biv02}; \L okas \& Mamon
\cite{lok03}). This picture is not supported by the analysis of
Ram\'{\i}rez \& de Souza (\cite{ram98}) who studied the deviations from
Gaussianity of the overall distribution of the line-of-sight
velocities of the galaxies. These authors concluded that the orbits of
ellipticals are close to radial, while spirals would have more
isotropic orbits.  However, van der Marel et al. (\cite{van00}) and Biviano
(\cite{biv02}) argue that the conclusion of Ram\'{\i}rez \& de Souza is most
likely due to erroneous assumptions in their modelling.

One of the most extensive dynamical analyses so far was done for 14
`regular' galaxy clusters from the CNOC (Carlberg et
al. \cite{car97b,car97c}).  Adopting {\em ad hoc} functional forms for
the 3-D number density, the mean squared components of the radial
velocity, and the velocity anisotropy profile, Carlberg et al.
(\cite{car97b,car97c}) concluded that the velocity anisotropy is zero
or at most mildly radial. The CNOC data were re-analysed by van der
Marel et al. (\cite{van00}), who used the method developed by van der
Marel (\cite{van94}), assuming a three-parameter family of
mass-density profiles, and a set of constant values for the velocity
anisotropy, to determine the parameters in the mass-profile model from
the best fit to the line-of-sight velocity dispersion profile. More
recently, from the analysis of the projected phase-space distribution
of $\sim 15,000$ galaxies in the infall regions of eight nearby
clusters (the CAIRNS project), Rines et al. (\cite{rin03}) concluded
that galaxy orbits are consistent with being isotropic within the
virial radius. Note that neither van der Marel et al. (\cite{van00}), nor
Rines et al. (\cite{rin03}) distinguished among different cluster
galaxy populations.

In this paper we study the galaxy orbits in an ensemble cluster of
3056 galaxy members of 59 clusters observed in the ENACS. We use the
`inversion' of the Jeans equation of stellar dynamics, as derived by
Binney \& Mamon (\cite{bin82}), and we apply the solution method given
by Solanes \& Salvador-Sol\'e (\cite{sol90}, hereafter S$^2$). The
analysis requires the mass profile $M(<r)$, for which we use the
estimate derived by Katgert et al. (\cite{kat04}, paper XII) for the
same ensemble cluster. Preliminary results were discussed by Biviano
et al.  (\cite{biv99,biv03,biv04}), Mazure et al. (\cite{maz00}),
Biviano (\cite{biv02}), and Biviano \& Katgert (\cite{bivk03}).

In \S~\ref{s-data} we summarize the data that we use, describe the
different classes of cluster galaxies, and the construction of the
`ensemble cluster'. In \S~\ref{s-profiles} we discuss the
observational basis for our analysis, i.e. the projected and
de-projected number density profiles, and the velocity-dispersion
profiles of the various galaxy classes. In \S~\ref{s-mass} we
summarize the observed mass profile and the model fits that we used in
the analysis of the orbits. In \S~\ref{s-method} we summarize the
inversion procedure by which we derived the velocity-anisotropy
profiles for the brightest ellipticals, the early spirals, the late
spirals, and the galaxies in substructures. In
\S~\ref{s-orbits} we describe the results of the analysis, and in
\S~\ref{s-disc} we discuss the implications of the results for
ideas about the evolution of the clusters themselves, and about the
formation and evolution of galaxies in clusters. Our summary and
conclusions are given in \S~\ref{s-summ}. Throughout this paper
we use $H_0 = 100 \, h$ km~s$^{-1}$~Mpc$^{-1}$.

\section{The data, the galaxy classes, and the ensemble cluster}
\label{s-data}

Our analysis of the orbits of galaxies in rich clusters is based on
data obtained in the context of the ENACS. Katgert et
al. (\cite{kat96,kat98}, papers I and V of this series, respectively)
describe the multi-object fiber spectroscopy with the 3.6-m telescope
at La Silla, as well as the photometry of the 5634 galaxies in 107
rich, nearby ($ z\la 0.1$) Abell clusters. After the spectroscopic
survey was done, a long-term programme of CCD-imaging with the Dutch
92-~cm telescope at La Silla was carried out which has yielded
photometrically calibrated images for 2295 ENACS galaxies. Thomas
(\cite{tho04}, paper VIII) has used those images to derive morphological
types, with which he also refined and recalibrated the galaxy
classification based on the ENACS spectra, as carried out previously
in paper VI.

The morphological types derived by Thomas were supplemented with
morphological types from the literature, and those were combined with
the recalibrated spectral types from paper VI into a single
classification scheme. This has yielded galaxy types for 4884 ENACS
galaxies, of which 56\% are purely morphological, 35\% are purely
spectroscopic, and 6\% are a combination of both. The remaining 3\%
had an early morphological type (E or S0) but showed emission lines in
the spectrum. With these galaxy types, Thomas \& Katgert
(\cite{thok04}, paper X) studied the morphology-radius and
morphology-density relations.  These galaxy types also form the basis
of the study of morphology and luminosity segregation (paper XI).

In paper XI the galaxy classes were defined that must be distinguished
because they have different phase-space distributions. In particular,
this applies to galaxies within and outside substructures. The
membership of a given galaxy to a substructure was determined using a
slightly modified version of the test of Dressler \& Shectman
(\cite{dre88}). In this test, a quantity $\delta$ was computed for
each galaxy, designed to indicate when the neighbourhood of the galaxy
is characterized by a different average velocity, and/or a smaller
velocity dispersion than the cluster mean values (see paper XI for
details). Galaxies with $\delta \leq 1.8$ were shown to have a very
small probability of belonging to substructures. On the other hand,
only two thirds of the galaxies with $\delta > 1.8$ really belong to
substructures. In the present paper, we use $\delta=1.8$ to separate
galaxies within substructures from galaxies outside substructures.
However, we also checked our results for the galaxies in substructures
with $\delta > 2.2$. Clearly, the $\delta > 2.2$ sample is smaller
than the $\delta > 1.8$ sample, but there is less contamination by
galaxies outside substructures. The results for the $\delta > 1.8$
sample are confirmed from the $\delta > 2.2$ sample. Therefore, for
the sake of simplicity, in the rest of this paper we only refer to
'galaxies in substructures' (or, more simply, 'Subs', in the
following), meaning galaxies with $\delta > 1.8$, keeping in mind that
the same results apply for the galaxies with $\delta > 2.2$.

In paper XI we showed that four classes of cluster galaxies must be
distinguished among the galaxies outside substructures, on the basis
of thei projected phase-space distributions. These are: (i)
the brightest ellipticals (with $M_R \leq -22+5 \log h$), which we
will refer to as `\ebr', (ii) the other ellipticals together with the
S0 galaxies (to be referred to as `Early'), (iii) the early spirals
(Sa--Sb), which we will denote by `\spe', and (iv) the late spirals
and irregulars (Sbc--Irr) together with the ELG (except those with
early morphology), or `\spl' for short.

Summarizing, we consider 5 classes of cluster galaxies: \ebr, Early,
\spe, \spl, and Subs, containing 34, 1129, 177, 328, and 686 galaxies,
respectively.  As explained in Appendix B.1 of paper XII, corrections
for incomplete azimuthal coverage in the spectroscopic observations
and sampling incompleteness had to be applied in the construction of
the number density profiles. In order to keep these correction factors
sufficiently small, galaxies located in poorly-sampled regions were
not used and those have not been included in the numbers given above.

The present analysis requires that data for several clusters are
combined into an ensemble cluster, to yield sufficient statistical
weight. If clusters form a homologous set, the ensemble cluster
effectively represents each of the clusters, provided that the correct
scaling was applied. Support for the assumption of homology comes from
the existence of a fundamental plane that relates some of the cluster
global properties (Schaeffer et al. \cite{sch93}; Adami et al. 
\cite{ada98b}, paper IV; Lanzoni et al. \cite{lan04}). As shown by 
Beisbart et al. (\cite{bei01}), clusters with substructure deviate
from that fundamental plane. Instead of eliminating all clusters with
signs of substructure, we have chosen to consider separately those
galaxies that are in substructures.

\begin{figure*}
\centering
\includegraphics[width=14cm]{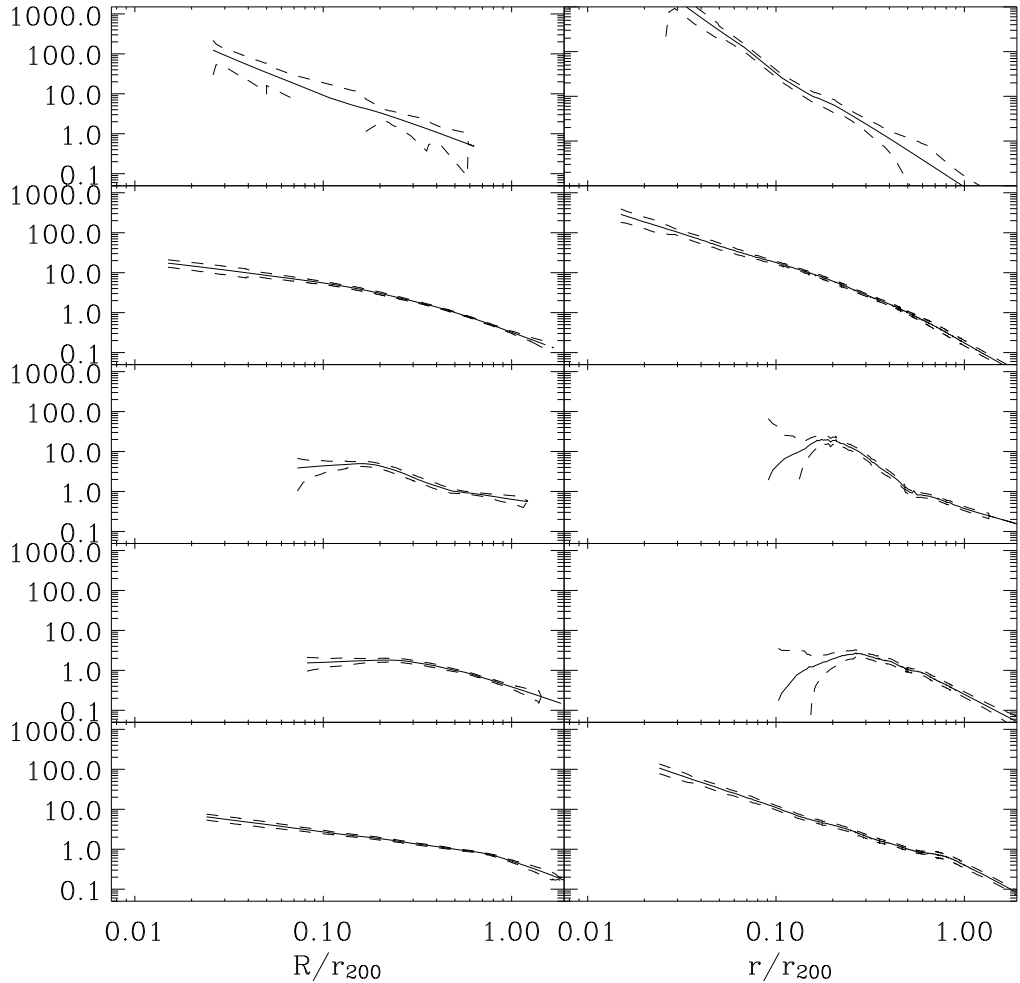}
\caption{Left: The best LOWESS estimate (solid line) of the projected 
number density $I(R)$, within the 1-$\sigma$ confidence interval
determined from bootstrap resamplings (dotted lines), for each of the
5 galaxy classes, from top to bottom:
\ebr, Early, \spe, \spl, Subs. The scale on the y-axis is
arbitrary. Right: Same as left panel, but for the de-projected number
density $\nu(r)$.}
\label{f-dprof}
\end{figure*}

As in papers XI and XII, we combined the data for 59 clusters with $z
< 0.1$, each with at least 20 member galaxies with ENACS redshifts,
and with galaxy types for at least 80\% of the members (see Table~A.1
in paper XI). The resulting ensemble cluster contains 3056 member
galaxies, for 2948 (or 96\%) of which a galaxy type is available. The
selection of cluster members was based on the method of den Hartog \&
Katgert (\cite{den96}), and its application to the ensemble cluster is
summarized in Appendix A of paper XII. We refer to papers XI and XII
for details on the way in which the data for many clusters were
combined. Those details concern the uniform method for the
determination of cluster centres, and the correct scaling of projected
distances from the cluster centres, $R$ (with $r_{200}$), and of
relative line-of-sight velocities (with the global line-of-sight
velocity dispersion $\sigma_p$). The scaling with $r_{200}$ ensures
that we avoid, as much as possible, mixing inner virialized cluster
regions with external non-virialized cluster regions. Note that the
scaling factors $r_{200}$ and $\sigma_p$ are computed using {\em all}
cluster members.

We assume that the ensemble cluster is spherically symmetric, not
rotating, and in a steady state. As discussed at length in Appendix C
of paper XII, these are reasonable assumptions for our ensemble
cluster.

\section{The number-density and velocity-dispersion profiles}
\label{s-profiles}

The observational basis for our study of the orbits of galaxies in
clusters is provided by the projected number-density profiles $I(R)$,
and the velocity-dispersion profiles $\sigma_p(R)$ for the 5 galaxy
classes that we consider, viz. `\ebr', `Early', `\spe', `\spl', and
'Subs' (see \S~\ref{s-data}). Here we summarize the steps involved in
the determination of these profiles, and their de-projection. Full
details can be found in Appendix~B of paper XII.

For the application of the Jeans equation -- to derive the mass
profile --, and its `inversion' -- to derive the velocity-anisotropy
profile --, smooth estimates of number density profiles,
velocity-dispersion profiles and combinations thereof are required. We
used the LOWESS technique (e.g. Gebhardt et al. \cite{geb94}) to
obtain smooth estimates of $I(R)$ and $\sigma_p(R)$. Whereas Gebhardt
et al. (\cite{geb94}) applied the LOWESS technique only to the
estimation of a velocity dispersion profile, we also developed a
variant that produces a smooth estimate of the number density profile.

The LOWESS technique yields estimates of $I(R)$ and $\sigma_p(R)$ at
the projected distance $R$ of each galaxy. These estimates are based
on a weighted linear fit to local estimates of projected density and
velocity dispersion. The linear fits typically involve between 30 and
80\% of the data points, but with a weight that drops steeply away
from the galaxy in question. The number density profiles, $I(R)$'s,
were corrected for sampling incompleteness, assuming axial symmetry.
Bootstrap resamplings yield estimates of the 68\% confidence limits
(approximately 1$\sigma$-errors) of the LOWESS estimate. The projected
number density profiles $I(R)$ of the 5 galaxy classes are shown in
the left-hand panels of Fig.~\ref{f-dprof}, together with their 68\%
confidence limits.

In the Jeans equation as well as in its `inversion' one also needs the
de-projected 3-D number density $\nu(r)$. In the right-hand panels of
Fig.~\ref{f-dprof} we show the $\nu(r)$-profiles, as derived by
de-projection via the Abel integral:
\begin{equation}
\nu(r) = - \frac{1}{\pi} \int_{r}^{\infty} \frac{{\rm d} I}{{\rm d} R} 
\frac{{\rm d} R}{\sqrt{R^2-r^2}} 
\label{e-abel1}
\end{equation}
This de-projection involves no assumptions other than spherical
symmetry, the extrapolation of $I(R)$ beyond the last measured point
towards large radii (for which we assume a tidal radius of 6.67
$r_{200}$), and continuity of $I(R)$ and its derivative at the last
measured point. We checked that the de-projected profiles are
essentially independent of the detailed form of the extrapolated
$I(R)$.

\begin{figure}
\centering
\includegraphics[width=9cm]{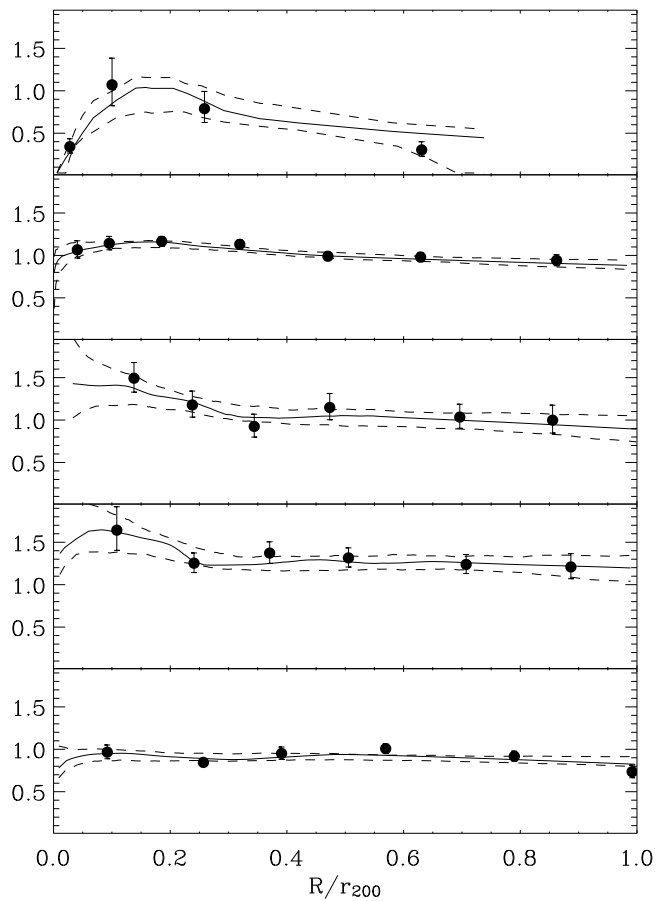}
\caption{The best LOWESS estimate (heavy line) of $\sigma_p(R)$,
together with the 68\% confidence levels (dashed lines), for each of
the 5 galaxy classes, from top to bottom: \ebr, Early, \spe,
\spl, Subs. The filled circles with error bars indicate binned
biweight estimates of $\sigma_p(R)$.}
\label{f-vdp}
\end{figure}

In Fig.~\ref{f-vdp} we show the projected velocity dispersion profiles
of the 5 galaxy classes as determined with the LOWESS technique. In
the same figure, we also show binned estimates of the velocity
dispersion, where the value of $\sigma_p(R)$ in each radial bin is
computed using the robust biweight estimator (see Beers et
al. \cite{bee90}).

To our knowledge, this is the first time that the number-density and
velocity-dispersion profiles for these 5 cluster galaxy classes have
been derived with such accuracy and in such detail. Therefore, we
briefly comment on the qualitative nature of the different $I(R)$,
$\nu(r)$ and $\sigma_p(R)$ before proceeding with the analysis.

Among galaxies outside substructure, the \ebr~ have the steepest
density profile in the centre, followed by the Early, the \spe, and
the \spl.  This is a clear manifestation of the morphology-density
relation (e.g. Dressler \cite{dre80}), and of luminosity segregation
(e.g. Rood \& Turnrose \cite{roo68} and paper XI). Interestingly, the density
profiles of both \spe~ and \spl~ decrease towards the cluster centre,
a clear indication that these galaxies avoid the central cluster
regions. On the contrary, \ebr~ are mostly found in the central
cluster regions. 

The Subs galaxies have a number-density profile that is rather steep
in the centre, but shows a weak 'plateau' at $\sim 0.6 \,
r_{200}$. Note that the number density profile of this galaxy class
could, in principle, be biased by systematic effects due to the
selection procedure of the members of substructures, which might
result in a radius-dependent detection efficiency. A comparison of the
de-projected number densities of the Subs-class galaxies and of the
bulk of the galaxies outside substructures, viz. the Early-class
galaxies (right-hand panels of Fig.~\ref{f-dprof}), shows that, within
$\sim 0.6 \, r_{200}$, the two profiles have essentially identical
logarithmic slopes. Beyond $\sim 0.6 \, r_{200}$ the number-density
profile of the Subs galaxies is quite a bit flatter than that of the
Early galaxies, until it steepens again beyond $\sim 1.0 \, r_{200}$.
This was already noted in paper XI. A comparison of the number-density
profile of the Subs galaxies with that obtained by De Lucia et
al. (\cite{del04}) from their numerical models of substructures in cold dark
matter haloes gives a similar result. The logarithmic slope between
$0.1 \, r_{200}$ and $0.8 \, r_{200}$ of the number-density of haloes
with masses $\sim 10^{13} M_{\odot}$ is about $-1.6$, not very
different from that of the Subs galaxies which is $-1.5$.

The velocity dispersion of the \ebr~ strongly decreases towards the
centre, with a slower but equally large decrease outwards (remember
that {\em all} velocity dispersions are normalized by the same, global
velocity dispersion calculated for {\em all} galaxies irrespective of
type). The special formation history and location of the \ebr~ at the
bottom of the cluster potential well is reflected in their very low
central velocity dispersion. In contrast, galaxies of the Early class
have a rather flat velocity-dispersion profile, changing by only
$\approx \pm 20$\% over the virial region. The velocity-dispersion
profiles of \spe~ and \spl~ are rather similar, starting at high
values near the centre with a fairly rapid decrease out to $r \approx
0.3 \, r_{200}$, and flattening towards larger projected distances.
Yet, the velocity dispersion of the \spl~ is larger than that of the
\spe~ (and, in fact, of any other class) at all radii. It is perhaps
interesting to note that the velocity-dispersion profiles of \spe~ and
\spl~ are remarkably similar to those of, respectively, the
`backsplash' and infalling populations of subhaloes found in the
numerical simulations of Gill et al. (\cite{gil04}).

Finally, the velocity-dispersion profile of the Subs class is very
'cold' and flat, even flatter and 'colder' than that of the Early
class. One might wonder if this is due to the procedure by which the
galaxies of the Subs class were selected, but it is very unlikely that
the velocity dispersion of the Subs class is biased low by the
selection. If anything, the actual velocity dispersion of the
subclusters is overestimated because the internal velocity dispersion
of the subclusters has not been corrected for. In \S~\ref{ss-subs} we
discuss several estimates for the real velocity-dispersion profile,
i.e. corrected for internal velocity dispersion and possible bias due
to the selection.

\section{The mass profile}
\label{s-mass}

In addition to the observed $I(R)$-, $\nu(r)$- and
$\sigma_p(R)$-profiles presented in \S~\ref{s-profiles} we also need
an estimate of the mass profile $M(<r)$ for a determination of the
$\beta(r)$-profiles. The mass profile that we will use here is the one
that was derived in paper XII, from the number density and
velocity-dispersion profiles of the Early-class galaxies. As discussed
in detail in paper XII, the Early-class galaxies are likely to be in
equilibrium with the cluster potential, as the formation of most of
them probably antedates their entry into the cluster, so that they
have had ample time to settle in the potential. In paper XII we also
showed that galaxies of the Early class have a nearly isotropic
velocity distribution; this follows from an analysis of the shape of
the distribution of their line-of-sight velocities. More specifically,
assuming a constant velocity anisotropy for the Early-class galaxies,
in paper XII we concluded that $-0.6 \la
\beta \la 0.1$, where
\begin{equation}
\beta(r) \equiv 1 - \frac{\rm{<} v_t^2 \rm{>}(r)}{\rm{<} v_r^2 \rm{>}(r)},
\label{e-beta}
\end{equation}
and $\rm{<} v_r^2 \rm{>}(r)$, $\rm{<} v_t^2 \rm{>}(r)$ are the mean
squared components of the radial and tangential velocity (see, e.g.,
Binney \& Tremaine \cite{bin87}). In this paper, we will often use the
parameter $\beta'$ instead of $\beta$ to describe the velocity
anisotropy, where $\beta'$ is defined as follows:
\begin{equation}
\beta' \equiv (\rm{<}
v_r^2 \rm{>}/\rm{<} v_t^2 \rm{>})^{1/2} \equiv (1-\beta)^{-1/2}.
\label{e-beta'}
\end{equation}
The constraint that we derived in paper XII for $\beta$ of the
Early-class galaxies translates into $\beta' \simeq
1.0_{-0.2}^{+0.05}$.

For an isotropic velocity distribution ($\beta' = 1.0$, or $\beta =
0$) the mass profile follows from the isotropic Jeans equation:
\begin{equation}
M(<r) = - \frac{r \rm{<} v_r^2 \rm{>}}{G} \left( \frac{{\rm d}
\ln \nu}{{\rm d} \ln r} + \frac{{\rm d} \ln \rm{<} v_r^2 \rm{>}}{{\rm d} 
\ln r} \right),
\label{e-jeans}
\end{equation}
where $\rm{<} v_r^2\rm{>}(r)$ follows from:
\begin{equation}
\rm{<} v_r^2 \rm{>}(r) = - \frac{1}{\pi \nu(r)} \int_{r}^{\infty}
\frac{{\rm d} [I(R) \times \sigma_p^2(R)]}{{\rm d} R} 
\frac{{\rm d} R}{\sqrt{R^2-r^2}}.
\label{e-abel3d}
\end{equation}

 As with the de-projection of $I(R)$, Eq.~\ref{e-abel3d} requires
extrapolation of $\sigma_p(R)$ to the tidal radius (for details, see
Appendix B.2 in paper XII).

The resulting $M(<r)$, and its derivative $\rho(r)$ are shown in
Fig.~4 of paper XII. They are very well represented by a NFW profile 
(Navarro et al. \cite{nav97}) with a scaling radius 
$r_s=0.25_{-0.10}^{+0.15} \, r_{200}$.

\section{The S$^2$ method for the solution of $\beta'(r)$}
\label{s-method}

Binney \& Mamon (\cite{bin82}) were the first to show that it is
possible to derive $\beta(r)$ when $I(R)$, $\sigma_p(R)$ and $M(<r)$
are known. S$^2$ gave a practical recipe for application of the
method, and we give a brief summary of their method to determine the
velocity-anisotropy profile $\beta(r)$ for a given class of galaxies
in equilibrium in a cluster gravitational potential with mass profile
$M(<r)$. Hereafter we give a brief summary of the S$^2$ method (note
that in this context we use $\beta$ instead of $\beta'$ to be
consistent with the earlier papers).

The estimate of the mass profile $M(<r)$ is used together with the
estimate of the 3-D number density $\nu(r)$ (derived from $I(R)$ as
before, see eq.~\ref{e-abel1}), to calculate $\Psi(r) = - G M(<r) \,
\nu(r) / r^2$. The observed functions $\sigma_p(R)$ and $I(R)$ are
used to derive $H(R) = {1 \over 2} \, I(R) \, \sigma^2_p(R)$, which in turn
is used to calculate the function $K(r)$ by the Abel integral:
\begin{equation}
K(r) = 2 \int_{r}^{\infty} H(x) \frac{x \, {\rm d} x }{\sqrt{x^2 - r^2}}.
\label{e-s3-kr}
\end{equation}
Using the functions $\Psi(r)$ and $K(r)$, one obtains the 
following two equations for $\rm{<} v_r^2 \rm{>} (r)$ and $\beta(r)$:
\begin{eqnarray}
[3 - 2 \beta(r)] \times \rm{<} v_r^2 \rm{>}(r) = &
	\frac{-1}{\nu(r)} \int_{r}^{\infty} \Psi(x) \, {\rm d} x - \\ 
	& - \frac{2}{\pi r \nu(r)} \frac{{\rm d} K(r)}{{\rm d} r} \nonumber
\label{e-s3-a}
\end{eqnarray}
and
\begin{eqnarray}
\label{e-s3-b}
\beta(r) \, \rm{<} v_r^2 \rm{>}(r) = & \frac{1}{\nu(r) r^3}
	\int_{0}^{r} x^3 \Psi(x) {\rm d} x + \\
        & + \frac{1}{\pi r \nu(r)} \frac{{\rm d} K(r)}{{\rm d} r} - 
	\nonumber \\
	& -\frac{3 K(r)}{\pi r^2 \nu(r)} +
	\frac{3}{\pi r^3 \nu(r)} \int_{0}^{r} K(x) {\rm d} x \nonumber
\end{eqnarray}
from which $\rm{<} v_r^2 \rm{>} (r)$ and $\beta(r)$ can be derived.

The practical application of the method is far from trivial. First,
one needs a smooth representation of the mass profile, which can be
extrapolated confidently to large radii where we have not measured it.
The extrapolation is done by using analytic mass profiles that
adequately fit the $M(<r)$, such as the NFW profile (see paper XII).
This ensures that the integral of $\Psi(r)$ in Eq.~\ref{e-s3-a} (whose
upper integration limit we set to $6.67 \, r_{200}$; see
\S~\ref{s-profiles}) is not problematic.  Fortunately, $\Psi(r)$
(which is negative) asymptotically approaches 0 with increasing $r$,
and it does so with a sufficiently flat slope that the exact choice of
the upper integration limit and the analytic representation of $M(<r)$
used for the extrapolation, do not influence the integral of $\Psi(r)$
in a significant way.

Secondly, one needs to extrapolate the observed velocity-dispersion
profiles, without having very strong constraints. For each class, we
check that different (plausible) extrapolations have no significant
effect on the results of the S$^2$ procedure within the observed
radial range.

A third important point is that Eq.~\ref{e-s3-b} contains two
integrals which have a lower integration limit of $r=0$. Because it is
quite difficult to determine the two integrands ($r^3 \Psi(r)$ and
$K(r)$) at very small $r$ from observations, a plausible interpolation
of $r^3 \Psi(r)$ and $K(r)$ from the innermost measured `point' to
$r=0$ (for which both $r^3 \Psi(r)$ and $K(r)$ are known from first
principles) is needed. We made a special effort to ensure plausible
interpolations from the innermost point for which the data is
available to $r=0$, using low-order polynomials.

It will not come as a surprise, given the equations involved, that it
is practically impossible to give estimates of the formal errors in
$\beta'(r)$ as derived with the S$^2$ method.  Approximate confidence
levels on the $\beta'(r)$ of each galaxy class were therefore
determined by estimating the r.m.s. of four $\beta'(r)$, obtained by
applying the S$^2$ method to four subsamples, each half the size of
the original sample. The fact that each subsample only contains half
the number of galaxies in the original sample, is likely to compensate
for the fact that the four subsamples are not all mutually
independent, which could lead to underestimation of the true
confidence levels.

\begin{figure}
\centering
\includegraphics[width=9cm]{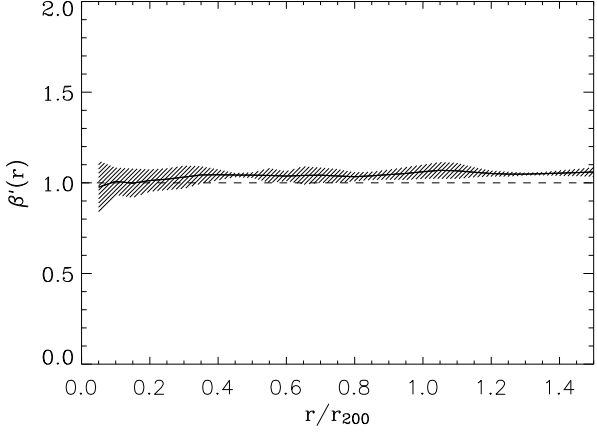}
\caption{The velocity anisotropy profile $\beta'(r) \equiv \rm{<} v_r^2 
\rm{>}^{1/2}/\rm{<} v_t^2 \rm{>}^{1/2}$, as 
derived for the galaxies of the Early class, using the
mass profile that was derived assuming that the same galaxies
have $\beta'(r) \equiv 1$. The shaded region indicates approximate
1-$\sigma$ confidence levels, as obtained by considering subsamples
half the size of the original sample. The value of $\beta'(r)$ 
is indeed quite close to 1, as it should.}
\label{f-beta0}
\end{figure}

We checked the robustness of our implementation of the S$^2$ method as
follows. We applied the S$^2$ method to the galaxies of the Early
class, adopting the mass profile that was determined using the same
galaxies as isotropic tracers (see paper XII and
\S~\ref{s-mass}). Clearly, one should obtain $\beta(r) \equiv 0$, or,
equivalently, $\beta'(r) \equiv 1$ (see eqs.~\ref{e-beta} and
\ref{e-beta'}). The result is shown in Fig.~\ref{f-beta0}. 
The shaded region indicates approximate 1-$\sigma$ confidence levels,
derived as described above. Indeed, we find a velocity anisotropy very
close to zero with $0.85 \leq \beta'(r) \leq 1.15$ over the radial
range $0 \leq r/r_{200} \leq 1.5$. Deviation from $\beta'(r) \equiv 1$
for the Early-class galaxies must be due to systematic errors arising
from extrapolation uncertainties, and numerical noise in the inversion
procedure (remember that our profiles are not analytic).  Yet, the
result in Fig.~\ref{f-beta0} indicates that our implementation of the
S$^2$ `inversion' works quite well.

We also applied a consistency test to all solutions that we obtained
with the S$^2$ method. I.e., we used the velocity-anisotropy profile
$\beta(r)$ obtained by the S$^2$ method for a given galaxy class, to
determine the projected velocity dispersion profile through
(see, e.g., van der Marel \cite{van94}):
\begin{equation}
\nu(r) \rm{<} v_r^2 \rm{>}(r) = - G \int_{r}^{\infty} \frac{\nu(\xi) \, 
M(<\xi)}{\xi^2} \, \exp \left[2 \int_{r}^{\xi} \frac{\beta \,
{\rm d} x}{x}\right] {\rm d} \xi
\label{e-vdm}
\end{equation}
and
\begin{equation}
I(R) \sigma_p^2(R) = 2 \int_{R}^{\infty} \left( 1 - \beta(r)\frac{R^2}{r^2}
\right) \frac{\nu \, r \, \rm{<} v_r^2 \rm{>}(r) \, {\rm d} r}{\sqrt{r^2-R^2}}
\label{e-abel2d}
\end{equation}
We then compared this predicted velocity-dispersion profile with the
observed $\sigma_p(R)$. In other words, we closed the loop, from
observables and the mass profile to $\beta(r)$, then from $\beta(r)$
and the mass profile back to the observables.

The observed and predicted $\sigma_p(R)$ are always in very good
agreement (see \S~\ref{s-orbits}), despite the fact that we cannot
determine $\beta(r)$ beyond $\sim 1.5 \, r_{200}$, while knowledge of
this function to very large radii is required to solve
eq.~\ref{e-abel2d}. The behaviour of $\beta(r)$ at large radii is not
important since the number-density profiles of all galaxy classes drop
sufficiently fast with radius. Even for the \spl, which have the
shallower $\nu(r)$, the effect of adopting two very different
extrapolations of $\beta(r)$ to large radii (one derived from the
analytical model proposed by \L okas \& Mamon \cite{lok01}, the other
from the numerical simulations of Diaferio \cite{dia99}) results in a
$\la 10$\% variation at any point of the predicted $\sigma_p(R)$.

\section{The velocity-anisotropy profiles}
\label{s-orbits}

We now investigate the orbits of the four classes of cluster galaxies
that were not used to determine the mass profile, viz. \ebr,
\spe, \spl, and Subs, in the gravitational potential 
determined using the galaxies of the Early class. In other words: we
try to construct equilibrium solutions for each of the galaxy classes,
with physically acceptable velocity-anisotropy profiles. However,
first we try to find solutions with isotropic orbits (or, $\beta'
\equiv 1$). For this we need to solve eq.~\ref{e-vdm}, using the
$\nu(r)$ of each class, and the mass profile $M(<r)$ as determined
using the Early-class galaxies, setting $\beta'(r) \equiv 1$. If the
comparison between the predicted and the observed velocity-dispersion
profile yields an acceptable $\chi^2$, we conclude that the galaxies
of the given class can be considered isotropic tracers of the cluster
gravitational potential. 

After trying the isotropic solution, we then use the S$^2$ method to
solve for $\beta'(r)$. Note that, unlike Carlberg et
al. (\cite{car97b,car97c}), van der Marel et al. (\cite{van00}), and
Rines et al. (\cite{rin03}) we do not prescribe a functional form for
$\beta'(r)$, nor do we assume a constant value for $\beta'(r)$.

\subsection{The brightest ellipticals}
\label{ss-ebr}

\begin{figure}
\centering
\includegraphics[width=9cm]{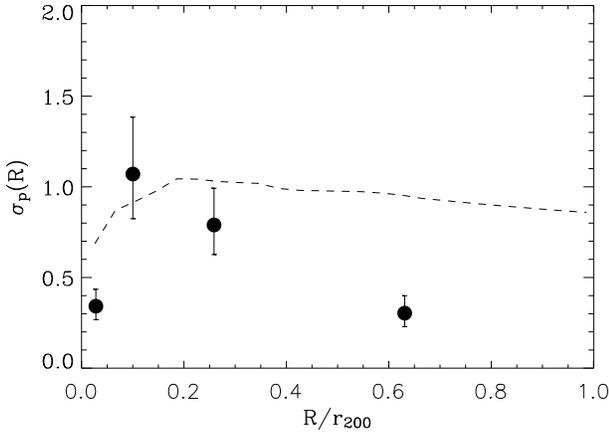}
\caption{The observed 
         velocity-dispersion profile $\sigma_p(R)$ of the \ebr~ galaxy
         class (dots with 1$\sigma$ errors), compared with the
         predicted $\sigma_p(R)$ (dashed line), assuming isotropic
         orbits in the gravitational potential determined from
         galaxies of the Early class.}
\label{f-ebr}
\end{figure}

The velocity-dispersion profile predicted for the \ebr~ class assuming
isotropic orbits is much flatter than the observed $\sigma_p(R)$ (see
Fig.~\ref{f-ebr}). We can reject the isotropic solution at $>99$\%
confidence level ($\chi^2=98$ on 4 data-points).

Interestingly, abandoning the isotropy assumption does not help. I.e.
there is no physical solution for which the \ebr~ are in equilibrium
in the cluster gravitational potential (i.e. the S$^2$ method predicts
negative $\rm{<} v_r^2\rm{>}$ and $\beta(r) > 1$ over most of the
radial range covered by our observations). There are two
straightforward interpretations of this result: either the galaxies of
the \ebr~ class are indeed out of dynamical equilibrium, or they do
not fulfil the conditions for the application of the Jeans
equation. We will return to this point in
\S~\ref{s-disc}.

\subsection{The early spirals}
\label{ss-spe}

\begin{figure}
\centering
\includegraphics[width=9cm]{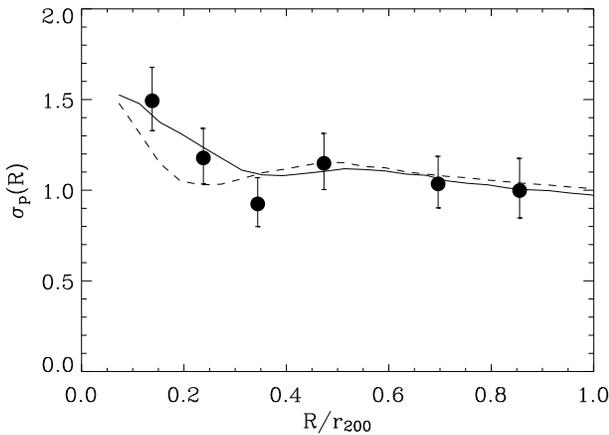}
\caption{The observed 
         velocity-dispersion profile $\sigma_p(R)$ of the \spe~ galaxy
         class (dots with 1$\sigma$ errors), compared with the
         predicted $\sigma_p(R)$ (dashed line), obtained by assuming
         isotropic orbits in the gravitational potential determined
         from galaxies of the Early class, and with the predicted
         $\sigma_p(R)$ (solid line), obtained by using the
         velocity-anisotropy profile $\beta'(r)$ determined by the
         S$^2$ method and shown in Fig.~\ref{f-betase}.}
\label{f-vdp-se}
\end{figure}

For the galaxies of the \spe~ class we do find acceptable equilibrium
solutions assuming an isotropic velocity distribution. The predicted
velocity dispersion profile provides an acceptable fit to the observed
$\sigma_p(R)$ ($\chi^2=5.2$ on 6 data-points, rejection probability of
61\%). This profile is shown in Fig.~\ref{f-vdp-se}, together with the
observations and their 1$\sigma$ errors. Note that, although the data
can be represented satisfactorily with isotropic orbits in the mass
profile determined using the Early-class galaxies, the innermost
values of $\sigma_p(R)$ are somewhat underpredicted.

\begin{figure}
\centering
\includegraphics[width=9cm]{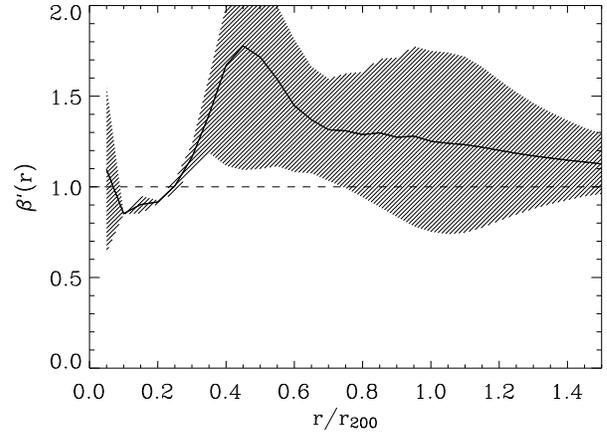}
\caption{The anisotropy profile, $\beta'(r) \equiv [\rm{<} v_r^2 
\rm{>}/\rm{<} v_t^2 \rm{>}]^{1/2}$ as 
derived for the galaxies of the \spe~ class, through the S$^2$
method. The shaded region indicates approximate 1-$\sigma$ confidence
levels, as obtained by considering subsamples half the size of the
original sample.}
\label{f-betase}
\end{figure}

The velocity-anisotropy profile of the \spe~ class (determined via the
S$^2$ method) is shown as the solid line in Fig.~\ref{f-betase}. The
velocity-anisotropy profile $\beta'(r)$ is very close to unity near
the centre, then rises to a maximum value of $\approx 1.8$ at $r
\approx 0.45 \, r_{200}$ and then decreases again to reach $\approx 1.1$
at $r/r_{200} \approx 1.5$. As mentioned before, we checked the
quality of this $\beta'(r)$ solution by calculating the implied
velocity-dispersion profile, solving eq.~\ref{e-vdm} for this
$\beta(r)$. The $\sigma_p(R)$ predicted in this way from the
$\beta'(r)$ indicated by the solid line in Fig.~\ref{f-betase}, is
shown in Fig.~\ref{f-vdp-se}, also as a solid line. As expected, the
latter is closer ($\chi^2=2.0$ on 6 data-points, rejection probability
of 16\%) to the observations than the isotropic solution (dashed line)
but not significantly so, because the isotropic model already yields an
acceptable fit to the data. As a matter of fact, the uncertainties on
the $\beta'(r)$ profile determined via the S$^2$ method are quite
large, so that any deviation from the isotropic solution is not really
significant.

\subsection{The late spirals+ELG}
\label{ss-spl}

\begin{figure}
\centering
\includegraphics[width=9cm]{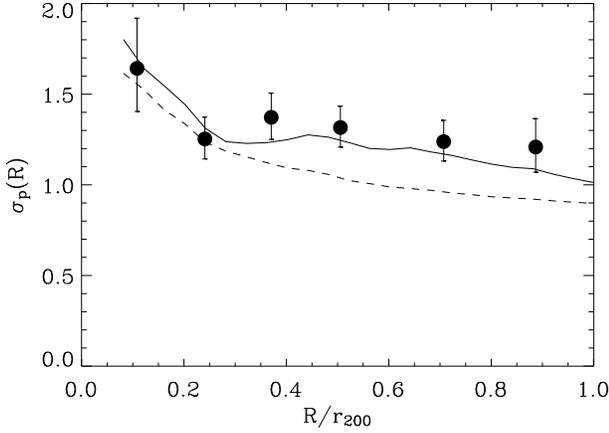}
\caption{The observed 
         velocity-dispersion profile $\sigma_p(R)$ of the \spl~ galaxy
         class (dots with 1$\sigma$ errors), compared with the
         predicted $\sigma_p(R)$ (dashed line), obtained by assuming
         isotropic orbits in the gravitational potential determined
         from galaxies of the Early class, and with the predicted
         $\sigma_p(R)$ (solid line), obtained by using the
         velocity-anisotropy profile $\beta'(r)$ determined by the
         S$^2$ method and shown in Fig.~\ref{f-betasl}.}
\label{f-vdp-sl}
\end{figure}

For the galaxies of the \spl~ class we do not find acceptable
equilibrium solutions assuming an isotropic velocity distribution.
This is illustrated in Fig.~\ref{f-vdp-sl}, where the predicted
$\sigma_p(R)$ (dashed line) is clearly seen to provide a poor fit to
the data ($\chi^2=18.2$ on 6 data-points, rejection probability
$>99$\%).  Beyond $R>0.3 \, r_{200}$ the predicted velocity-dispersion
profile is well below the observed values. Hence, purely isotropic
orbits are rejected.

\begin{figure}
\centering
\includegraphics[width=9cm]{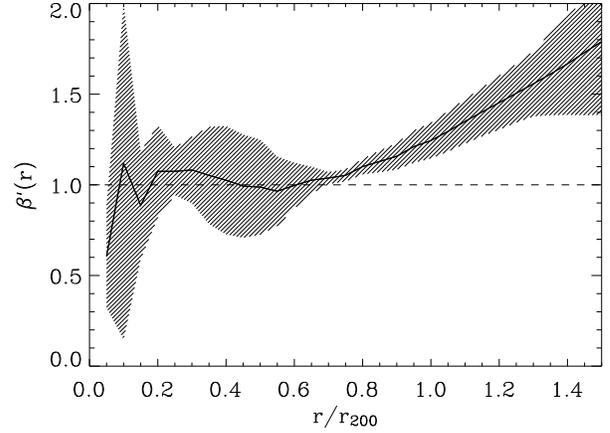}
\caption{The anisotropy profile, $\beta'(r) \equiv [\rm{<} v_r^2 
\rm{>}/\rm{<} v_t^2 \rm{>}]^{1/2}$ as 
derived for the galaxies of the \spl~ class, through the S$^2$
method. The shaded region indicates approximate 1-$\sigma$ confidence
levels, as obtained by considering subsamples half the size of the
original sample.}
\label{f-betasl}
\end{figure}

We then considered anisotropic solutions.  The velocity-anisotropy
profile of the \spl~ class (determined via the S$^2$ method) is shown
in Fig.~\ref{f-betasl}. The profile is very close to unity out to $r
\approx 0.7 r_{200}$, where it starts growing almost linearly with
radius to reach a value of $\approx 1.8$ at $r/r_{200} \approx
1.5$. As usual, we checked the quality of the
$\beta'(r)$-solution by calculating the implied velocity-dispersion
profile, solving eq.~\ref{e-vdm} for this $\beta(r)$. The
$\sigma_p(R)$ predicted in this way from the velocity-anisotropy
profile indicated in Fig.~\ref{f-betasl}, is shown in
Fig.~\ref{f-vdp-sl} as a solid line. As expected, it reproduces quite
well the observed $\sigma_p(R)$ of \spl.

In the case of the \spl~ class the velocity-dispersion profile
predicted with the $\beta'(r)$ obtained with the S$^2$ method not only
fits the data better than the isotropic case (this is also true for
the \spe-class galaxies), but it also does so in a significantly
better manner ($\chi^2=2.6$ on 6 data-points, rejection probability
23\%). Therefore, mild radial anisotropy is needed in order to put the
\spl-class galaxies in dynamical equlibrium in the cluster potential.

\subsection{The galaxies in substructures}
\label{ss-subs}
\begin{figure}
\centering
\includegraphics[width=9cm]{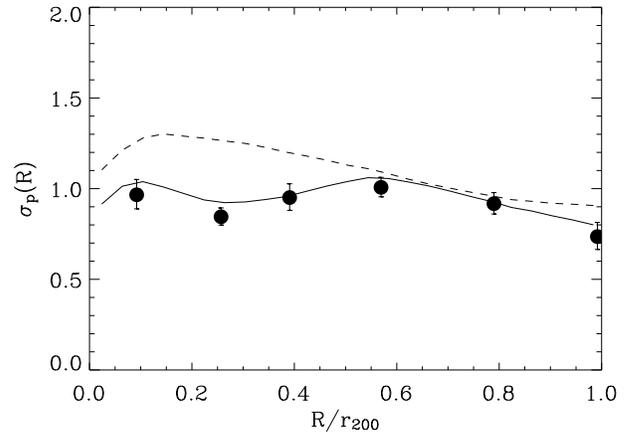}
\caption{The observed 
         velocity-dispersion profile $\sigma_p(R)$ of the Subs galaxy
         class (dots with 1$\sigma$ errors), compared with the
         predicted $\sigma_p(R)$ (dashed line), obtained by assuming
         isotropic orbits in the gravitational potential determined
         from galaxies of the Early class, and with the predicted
         $\sigma_p(R)$ (solid line), obtained by using the
         velocity-anisotropy profile $\beta'(r)$ determined by the
         S$^2$ method and shown by the solid line in Fig.~\ref{f-betasubs}
         (lower panel).}
\label{f-vdp-subs}
\end{figure}

As for the \spl-class galaxies, we do not find acceptable equilibrium
solutions for the galaxies of the Subs class if we assume an isotropic
velocity distribution. As can be seen in Fig.~\ref{f-vdp-subs}, the
predicted $\sigma_p(R)$ (dashed line) is way off the data (dots with
error-bars; $\chi^2=118.5$ on 6 data-points, rejection probability
$>99$\%), and overestimates the observed velocity dispersion at
essentially all radii.

\begin{figure}
\centering
\includegraphics[width=9cm]{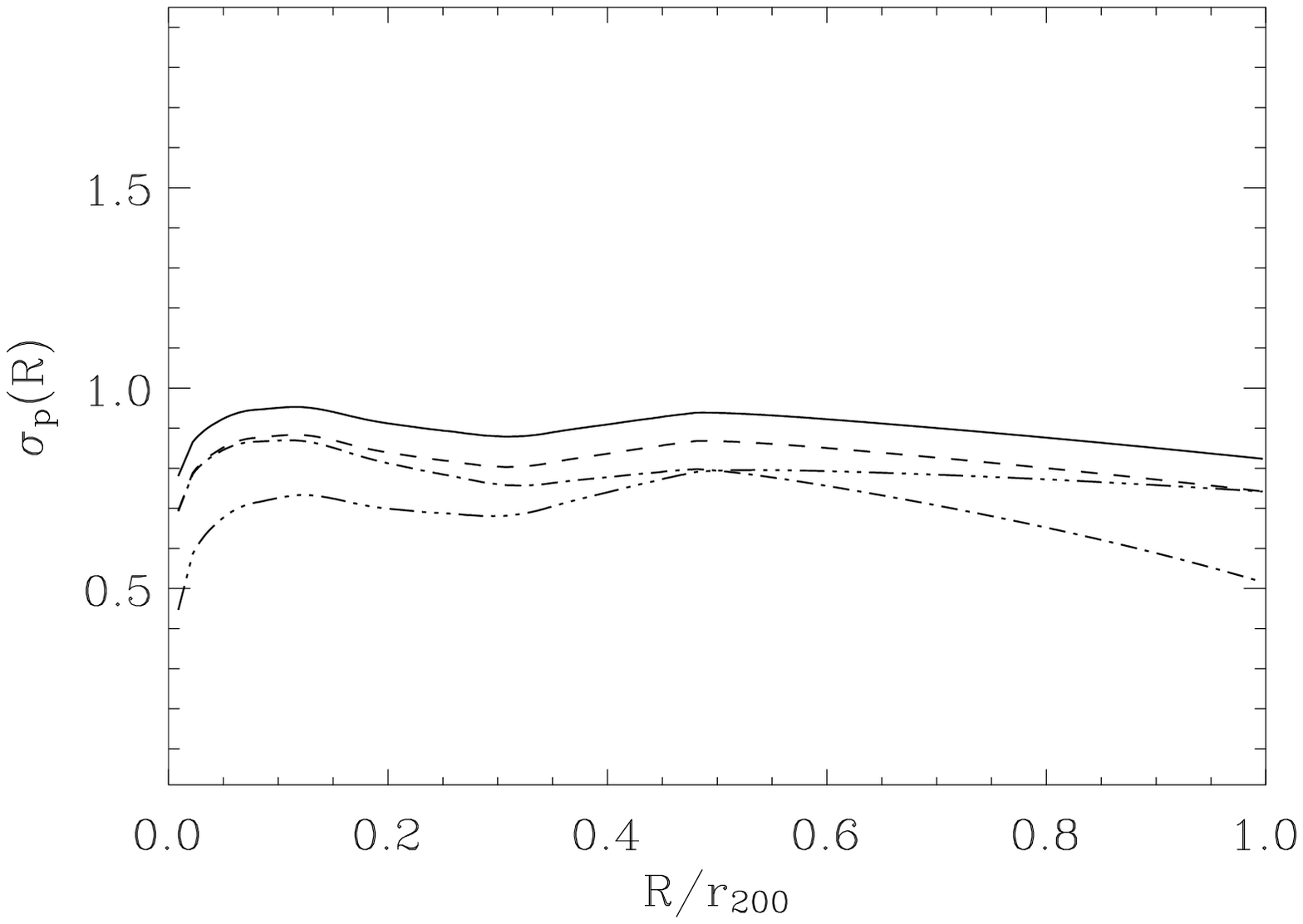}
\includegraphics[width=9cm]{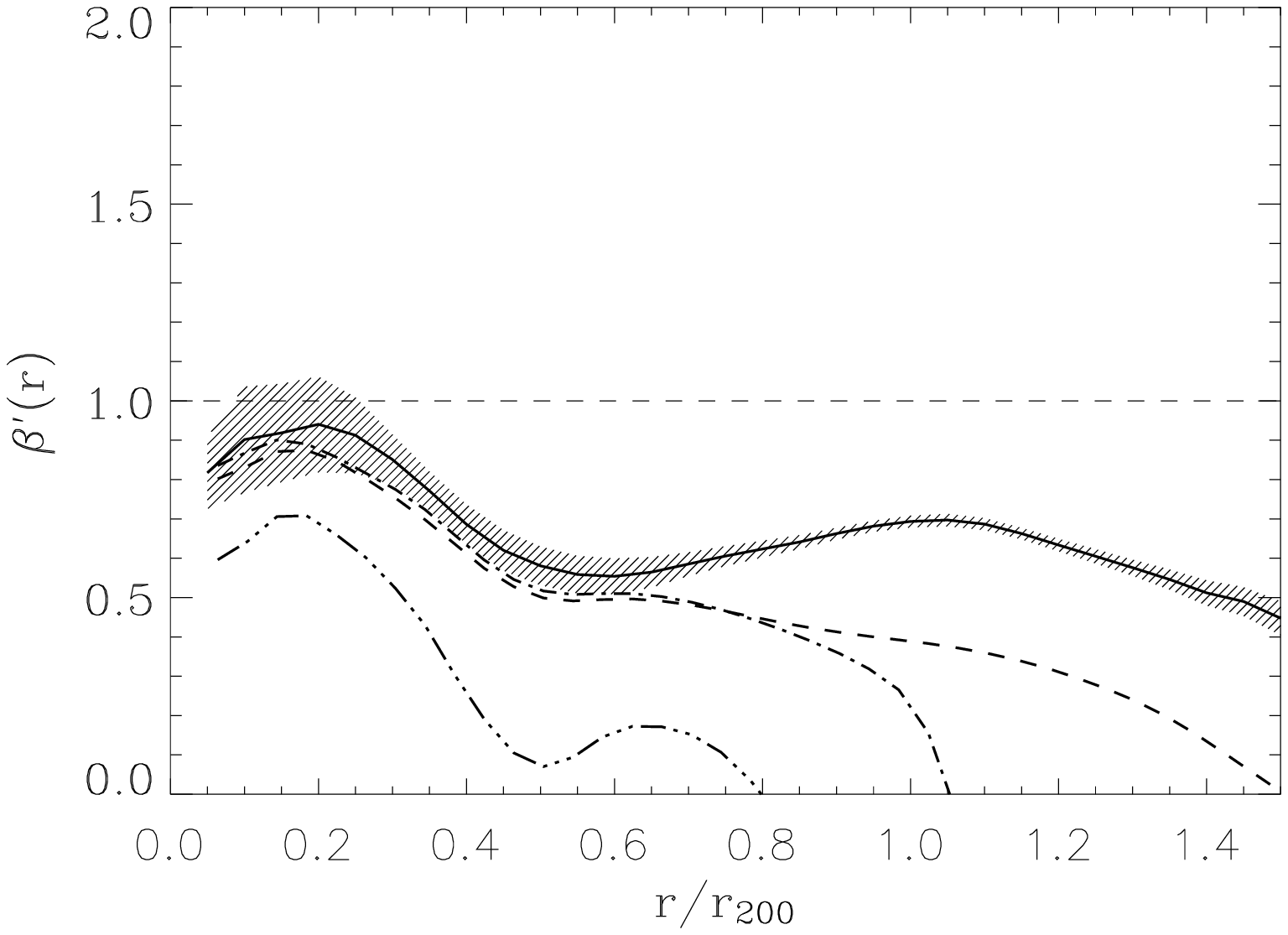}
\caption{Upper panel: the observed velocity-dispersion profile 
$\sigma_p(R)$ of the galaxies in subclusters (solid line). The dashed
line shows the result of deconvolving $\sigma_p(R)$ with an assumed
constant internal velocity dispersion of 250 km~s$^{-1}$. The two other
curves assume radius-dependent internal velocity dispersions in the
range 250--450 km~s$^{-1}$. Lower panel: the anisotropy profiles, $\beta'(r)
\equiv [\rm{<} v_r^2 \rm{>}/\rm{<} v_t^2 \rm{>}]^{1/2}$ derived for 
the galaxies of the Subs class, through the S$^2$ method with the
observed $I(R)$ and the four $\sigma_p(R)$-estimates shown in the upper
panel (with identical coding). The shaded region indicates approximate
1-$\sigma$ confidence levels around the solution that uses the
observed $\sigma_p(R)$-curve. These were obtained by considering
subsamples half the size of the original sample.}
\label{f-betasubs}
\end{figure}

Using the S$^2$ method for the Subs class,with the observed
$\sigma_p(R)$ we obtain the $\beta'(r)$ displayed as a solid line in
Fig.~\ref{f-betasubs}. The orbits are tangentially anisotropic at all
radii. As usual, we checked the $\beta'(r)$ solution in the space of
observables; the predicted $\sigma_p(R)$ is in excellent agreement
with the observed one (see Fig.~\ref{f-vdp-subs}; $\chi^2=5.1$ on 6
data-points, rejection probability $60$\%).

In the lower panel of Fig.~\ref{f-betasubs}, the shaded region
indicates approximate 1-$\sigma$ confidence interval, obtained as
described before. However, in this case the real confidence interval
is probably significantly larger, for two reasons.

First, in using the observed velocity-dispersion profile of the
galaxies in subclusters, we have ignored the internal velocity
dispersion of the subclusters. This means that the real velocity
dispersion is smaller than the observed one. We will make several
assumptions for the (possibly R-dependent) value of the apparent
internal velocity dispersion of the subclusters. In paper XI we
estimated the internal velocity dispersion of the identified
subclusters, and obtained a value of $\sim 400$--500 km~s$^{-1}$,
essentially independent of projected radius $R$. However, the true
internal velocity dispersion of a subcluster is likely to be smaller,
because the above estimate is biased high by galaxies that do not
belong to the subcluster but have been wrongly assigned to it by the
selection algorithm.

A more realistic estimate of a subcluster internal velocity dispersion
is probably 250~km~s$^{-1}$, a value close to the average velocity
dispersion of galaxy groups (see, e.g., Ramella et
al. \cite{ram89}). First, this constant value was subtracted in
quadrature from the observed $\sigma_p(R)$ of the Subs class to
produce the corrected velocity dispersion of subclusters shown as a
dashed line in the upper panel of Fig.~\ref{f-betasubs}. From this
corrected $\sigma_p(R)$, and using -- as before -- the observed
$I(R)$, we derived the corrected version of $\beta'(r)$, indicated by
the dashed line in the lower panel of Fig.~\ref{f-betasubs}). This
second solution implies even stronger tangential anisotropy of the
velocity distribution, which is not surprising since a larger fraction
of the (smaller) line-of-sight velocities is required to balance the
same cluster potential.

However, it is possible that due to the selection procedure, or for
physical reasons, we should not substract a {\em constant} value for
the internal velocity dispersion of the subclusters. Therefore, we
have assumed (rather arbitrarily) two different alternative solutions
for the real velocity-dispersion profile of the Subs galaxies. These
are shown as the dashed-dotted and dashed-triple-dotted curves in the
upper panel of Fig.~\ref{f-betasubs}. The former assumes a larger bias
in the observed velocity dispersion in the outer regions, while the
latter mimics a larger bias in the central region. The important point
in both assumptions is that the we must always deconconvolve the
observed $\sigma_p(R)$ with at least 250 km~s$^{-1}$ internal dispersion of
the subclusters.

For both assumptions about the real velocity-dispersion profile of the
Subs galaxies, we calculated $\beta'(r)$, assuming - as before -- that
the observed $I(R)$ is unbiased. The results are shown in the lower
panel of Fig.~\ref{f-betasubs}, where the same coding is used as in
the upper panel. Not surprisingly, the evidence for tangential
anisotropy of the Subs galaxies does not go away; if anything it gets
stronger (in the most extreme cases, no physical solution can be find
beyond a certain radius).

However, before we can accept this conclusion to be robust, we must
investigate the effect of possible biases in $I(R)$. Without real
modelling, we have considered two fairly extreme possibilities. In the
first one we assumed that the plateau around $0.6 \, r_{200}$ in the
observed $\nu(r)$ of the Subs galaxies (see Fig.~\ref{f-dprof} and
\S~\ref{s-profiles}) is (at least partly) an artefact due to the 
selection procedure. Consequently, we multiplied $\nu(r)$ by a
'constant' factor of 2 below $\sim 0.6 \, r_{200}$. The other
possibility assumes that the logarithmic slope of $\nu(r)$ below $0.6
\, r_{200}$ is $\simeq 40$~\% flatter than actually observed.

These two extreme assumptions about $\nu(r)$ were combined with the
various assumed estimates of $\sigma_p(R)$ to solve for $\beta'(r)$ of
the Subs galaxies. It appears that the conclusion of tangential orbits
is not affected by the different assumed shapes of $\nu(r)$ below $0.6
\, r_{200}$, and is thus primarily driven by the low values of the 
velocity dispersion of the Subs galaxies. In other words, we find
that, even without a detailed modelling of the selection effects and a
possible dependence of the internal velocity dispersion of the
subclusters on radius, the conclusion about the tangential orbits is
robust. The implications of this result will be discussed in
\S~\ref{s-disc}.

\section{Discussion}
\label{s-disc}

Adopting the mass profile as determined from the galaxies of the Early
class, we searched for equilibrium solutions for the other four
classes. As a first step, we assumed isotropy of the velocity
distribution, but subsequently we also solved for the anisotropy
profile $\beta'(r)$ using the S$^2$ set of equations (see
\S~\ref{s-method}). For the \ebr~class, we could not obtain
equilibrium solutions, no matter what we assumed for their
velocity-anisotropy profile. For the \spe~class the isotropic solution
was found to be quite acceptable. Yet, the velocity-anisotropy profile
of the \spe~class, as determined with the S$^2$-method, shows a slight
radial anisotropy at $r \approx 0.45 \, r_{200}$.  For the \spl~class,
the isotropic solution is rejected. Their velocity-anisotropy profile,
determined with the S$^2$ method, is close to zero out to $r
\approx 0.7 \, r_{200}$, and then increases almost linearly outwards,
reaching a radial anisotropy of $\beta' \approx 1.8$ (corresponding to
$\beta \approx 0.7$) at $r \approx 1.5 \, r_{200}$.  For the Subs
class the isotropic solution must also be rejected. Taken at face
value, the data for this class imply substantial tangential
anisotropy. However, this result may be affected by systematic effects
related to the selection of Subs galaxies. Until these effects have
been modelled in detail, the conclusion of tangential anisotropy must
be considered with some caution. 

Our conclusion that both early-type and late-type galaxies are in
equilibrium in the cluster potential, with the latter on more
radially-elongated orbits, is supported by several other studies in
the literature. The larger velocity dispersion and/or the
steeper velocity-dispersion profile of late-type galaxies with respect
to early-type galaxies, have often been interpreted as evidence for
infalling motions, and even for departure from virial equilibrium
(Moss \& Dickens \cite{mos77}; Sodr\'e et al.\cite{sod89}; paper III;
Adami et al. \cite{ada98a}; Solanes et al. \cite{sol01}). However,
Carlberg et al. (\cite{car97a}) already pointed out that the latter does not
need to be the case.  They found that red and blue galaxies in the
CNOC clusters are both in dynamical equilibrium in the cluster
gravitational potential, but they were not able to constrain the
velocity anisotropy of these galaxies.

From a more detailed analysis of the same dataset, van der Marel et
al. (\cite{van00}) were able to constrain the average velocity anisotropy
(assumed to be constant) of all CNOC cluster galaxies, to $0.75 \leq
\beta' \leq 1.2$ (95.4\% c.l.). This is a similar to what we found 
for the Early galaxies in paper XII, and from which we concluded that
those have isotropic orbits. Mahdavi et al. (\cite{mah99}) showed that
ELG in groups have an anisotropic velocity distribution, at the 95.4\%
c.l., with a best-fit constant $\beta' \approx 1.8$, whereas
absorption-line galaxies in groups have a best-fit constant $\beta'
\approx 1.4$, which is not, however, significantly different from
unity. Both values seem somewhat higher than the values we find, which
could indicate that the fraction of infalling galaxies is larger in
groups than in clusters.

Biviano (\cite{biv02}), also using the ENACS dataset, concluded that,
if absorption-line galaxies have zero anisotropy $\beta' = 1.0$, ELG
have an average constant anisotropy of $1.3 \leq \beta' \leq 1.6$
(68\% c.l.). This range is in reasonable agreement with our result for
the velocity anisotropy profile of the \spl~ galaxies (see
Fig.~\ref{f-betasl}), considering that many \spl~ galaxies are found
at large radii, where their radial velocity-anisotropy is largest.
Natarajan \& Kneib (\cite{nat96}) concluded that galaxies in A2218
have tangential orbits in the central region, and radial outside, with
an anisotropy profile resembling the one we find for \spl.  Finally,
Ram\'{\i}rez \& de Souza (\cite{ram98}) and Ram\'{\i}rez et
al. (\cite{ram00}) concluded that early-type galaxies have more
eccentric orbits than late-type galaxies, but their result arises from
incorrect assumptions in their method, as discussed by van der Marel
et al. (\cite{van00}) and Biviano (\cite{biv02}).

The present analysis is the first to consider the orbits of 5 distinct
classes of cluster galaxies. Using the Early class as a reference, we
find that 3 of the remaining 4 are in dynamical equilibrium within the
cluster gravitational potential; this is manifestly not the case for
the \ebr. The most likely explanation for our failure to find
solutions of the {\em collisionless} Jeans equation for the \ebr, is
that the \ebr~ either formed very near the cluster centre, or moved
there by losing kinetic energy subject to dynamical friction. At the
same time, they probably have grown through merging with other
galaxies. These processes lead to a loss of the orbital energy of
these galaxies. As a matter of fact, the very low velocity dispersion
of the \ebr~ at the cluster centre can be understood with the model of
Menci \& Fusco-Femiano (\cite{men96}), which is a solution of the {\em
collisional} Boltzmann-Liouville equation, and hence accounts for
galaxy collisions and merging processes.

The conclusion that the Early-class galaxies have a nearly isotropic
velocity distribution is not surprising, given the large body of
evidence indicating that ellipticals are an old cluster component. If
they form and become part of the cluster before it virializes, they
can obtain isotropic orbits through violent relaxation. From the
distribution of the ratio $r_{peri}/r_{apo}$ of the dark matter halos
in their simulations of rich clusters, Ghigna et al. (\cite{ghi98}) concluded
that about 25\% of the halos are on orbits more radial than 1:10,
where the median ratio is 1:6. Comparison with our result is not
immediate, but of the galaxies outside substructures 36\% belong to
the \spe~ and \spl~ classes. Since about two-thirds of those are late
spirals or ELG, which are the galaxies showing most of the velocity
anisotropy, these could indeed correspond to the halos with orbits
more radial than 1:10 in the $r_{peri}/r_{apo}$-ratio.

The increase of the radial-velocity anisotropy with radius of the \spl~
(see Fig.~\ref{f-betasl}) is a feature commonly found in numerical
simulations of dark matter haloes. E.g., the numerical simulations of
Tormen et al. (\cite{tor97}) predict an increasing radial velocity anisotropy
from $r/r_{200} \sim 0.3$ outwards, reaching $\beta' \sim 1.8$ at
$r/r_{200} \sim 1.5$, and the numerical simulations described by
Diaferio et al. (\cite{dia01}) predict a similar, though somewhat more
irregular, behaviour of the velocity anisotropy profile, with a
maximum anisotropy of $\beta' \sim 1.4$. This anisotropy profile
results from infall motions of the field haloes into the cluster, and
from the subsequent isotropization of the velocity distribution of
these haloes as they move towards the denser cluster centre.

The similarity of the $\beta'(r)$ of the \spl-class galaxies and of
the dark matter haloes in the models is quite interesting and it
probably means that the \spl~ galaxies still retain memory of their
infall motion from the field. The fact that a large fraction of the
\spl~ have emission-lines indicates that they have not yet lost their
gas as a consequence of tidal stripping, galaxy collisions, or ram
pressure. Hence it is unlikely that the \spl~ we observe have spent
much time in the hostile cluster environment, and many of them could
indeed even be on their first cluster crossing. 

On the other hand, the small velocity anisotropy of \spl~ near the
centre probably reflects the fact that the galaxies we identify as
\spl~ must avoid, or have avoided, the central region. Those \spl~ 
that have a significant radial anisotropy near the centre will cross
the very dense central cluster regions, where they cannot survive and
get disrupted, either to form dwarfs, or to contribute to the diffuse
intra-cluster light (Moore et al. \cite{moo99}). As a matter of fact,
\spl~ are not found in the cluster central regions (see
Fig.~\ref{f-dprof} and paper XI). The existence of faint spiral
structures in some dwarf spheroidals has now been demonstrated (Jerjen
et al. \cite{jer00}; Barazza et al. \cite{bar02}; Graham et
al. \cite{gra03}). Interestingly, Conselice et al. (\cite{con01}) found
that, in the Virgo cluster, the dwarf spheroidals have a velocity
distribution more similar to that of the spirals, than to that of the
ellipticals.

Our analysis shows that an acceptable equilibrium solution exists for
\spe~ with zero velocity anisotropy. Hence, it is likely that these 
galaxies are not very recent arrivals, since there is no evidence for
memory of their initial infall motion. These \spe~ galaxies are more
likely to survive the hostile cluster environment than \spl, because
of their higher surface brightness (see paper X; Moore et
al. \cite{moo99}).  This is consistent with the results of numerical
simulations showing that clusters contain red disk galaxies that,
after accretion from the field, attain dynamical equilibrium in $\sim
1$--2 Gyr (Diaferio et al. \cite{dia01}).

Additional indirect support for the scenario described above comes
from the similarity of the \spe~ and the \spl~ velocity dispersion
profiles with those of, respectively, the `backsplash' subhaloes, and
the subhaloes on first infall, identified by Gill et
al. (\cite{gil04}) in their N-body simulations of galaxy clusters.
This similarity suggests that many \spl~ could be on first infall,
while the \spe~ at large radii have already crossed the cluster core.

In paper X it was argued that \spe~ are likely to be the progenitors
of S0s. This conclusion is based on three different pieces of
evidence: (1) the strong increase of S0s in clusters since $z \sim
0.5$ (Dressler et al. \cite{dre97}; Fasano et al. \cite{fas00}),
accompanied by a similar decrease of the spiral fraction; (2) the
morphology-density relation (Thomas et al. \cite{thoh04}, hereafter
paper IX) which shows that the local projected density around \spe~ is
smaller than around S0s; and, (3) the strong similarity of the bulge
luminosity of \spe~ and S0s (paper X). If \spe~ transform into S0s and
if the velocity distribution of S0s is isotropic (see paper XII), it
is only natural that the velocity distribution of \spe~ is also
isotropic. Otherwise, the timescale of the
morphological-transformation process should be similar to that of the
velocity isotropization.

Even if the isotropic solution is perfectly acceptable for the \spe,
the data, when taken at face value, imply some radial velocity
anisotropy. Although the significance of the anisotropy is rather low,
we are tempted to speculate about a possible cause, if the anisotropy
at $r/r_{200} \approx 0.45$ were real. Galaxies with radial
velocity-anisotropy, moving on radially elongated orbits, will move
relatively fast near the cluster centre. It is possible, if not
likely, that the high radial velocity is a necessary condition for
\spe~ to avoid impulsive encounters and thus transformation into S0s 
in the central high-density cluster region.

Finally, galaxies in substructures provide an intrigueing view into
the processes that are important in the formation of clusters.
Recently, several groups have studied the properties of substructures
within dark-matter haloes over a range of total masses that includes
those of rich clusters (e.g. De Lucia et al. \cite{del04}; Taylor \& Babul
\cite{tay04}). We already mentioned the agreement between our radial
number-density profile of Subs galaxies and the radial distribution of
$10^{13} M_{\odot}$ substructures in the models of De Lucia et
al. (\cite{del04}, see \S~\ref{s-profiles}). Those models also show
that the more massive substructures are preferentially located in the
external regions of their parent haloes. This is most likely due to
tidal truncation and stripping of substructures that reach the dense
central regions. In addition, orbital decay can also contribute to
this mass segregation (e.g. Tormen et al. \cite{tor98}). The apparent
paucity of Subs galaxies in the inner regions of our clusters may thus
well be the result of mass segregation, instead of selection bias.

Taylor \& Babul (\cite{tay04}) discuss the evolution of the orbits of
the infalling substructures, and they conclude that disruption occurs
sooner for more radial orbits. This will lead to a tangentially
anisotropic distribution of orbits of the surviving substructures,
which is exactly what we find. So, even if the details of the
tangential anisotropy of the Subs galaxies requires additional
modelling, the result itself appears robust and not unexpected or
implausible.

\section{Summary and conclusions}
\label{s-summ}

We determined the equilibrium solutions for galaxies of the 4 classes
that were not used as tracers of the cluster potential. For this, we
solved the inverse Jeans equation, using the method of S$^2$.  We
found equilibrium solutions for galaxies of the \spe, the \spl, as
well as the Subs classes, but not for galaxies of the \ebr-class. The
equilibrium solution found for galaxies of the
\spe~ class was found to be consistent with them being on
isotropic orbits, except perhaps just outside the cluster central
region. On the other hand, isotropic solutions were found not to be
acceptable for galaxies of either the \spl~ or the Subs
classes. Galaxies of the \spl~ class were found to be on mildly radial
orbits, with the radial velocity-anisotropy increasing outwards. On
the contrary, tangential orbits seem to characterize galaxies of the
Subs class, but the significance of this result is difficult to assess
in view of possible systematics effects we have considered.

Our results support hierarchical models for the build-up of galaxy
clusters (see also paper XII).  Our results also constrain the
evolutionary history of cluster galaxies. They are consistent with, if
not suggestive of, a scenario where the very bright ellipticals form
very early, and sink to the bottom of the still forming cluster
potential well, losing orbital energy. In our scenario, the less
bright ellipticals, together with the S0s (the Early-class galaxies),
were already part of the cluster at the epoch of its formation, and
developed isotropic orbits through the process of violent relaxation,
or have lived sufficiently long in the cluster to have lost any memory
of original radial infall motions, through isotropization of their
orbits.

This is probably also the case for the early spirals, which make them
acceptable candidates for being the progenitors of S0s, also in view
of their structural properties. We speculate that some early spirals
near the cluster centre have managed to escape transformation into S0s
as a result of a selection effect in the velocity distribution.
Finally, many late spirals and emission-line galaxies (excluding those
of early morphology) are likely to be field galaxies recently arrived
into the cluster. Their radial infall motions are gradually
isotropized as they approach the cluster centre, until they get
disrupted or transformed into dwarf spheroidals as a consequence of
collisions and, in particular, tidal effects.

The galaxies in substructures apparently avoid the central regions and
they appear to be on tangential orbits. Although some modelling
remains to be done to assess the details of the implied anisotropy
profile, the conclusion of tangential anisotropy appears to be robust.
Interestingly, both effects are also seen in numerical simulations,
and they result from the mechanisms that 'destroy' the substructures
as they get nearer to the cluster cores.

\begin{acknowledgements}
We thank Gary Mamon, Alain Mazure, and Tom Thomas for useful
discussions. AB acknowledges the hospitality of Leiden Observatory. PK
acknowledges the hospitality of Trieste Observatory. This research was
partially supported by the Italian Ministry of Education, University,
and Research (MIUR grant COFIN2001028932 "Clusters and groups of
galaxies, the interplay of dark and baryonic matter"), and by the
Leids Kerkhoven-Bosscha Fonds.
\end{acknowledgements}

\vfill

\end{document}